\documentclass[twocolumn,showpacs,superscriptaddress,pra]{revtex4}

\usepackage{graphicx}
\usepackage{bm}

\begin{document}
\title{Mechanochemical coupling of kinesin studied with the neck linker swing model}
\author{Yaogen Shu}
\email{yaogen@itp.ac.cn} \affiliation{Institute of Theoretical
Physics, The Chinese Academy of Sciences,
 P. O. Box 2735, Beijing 100080, China}
\author{Hualin Shi}
\affiliation{Institute of Theoretical Physics, The Chinese Academy
of Sciences, P. O. Box 2735, Beijing 100080, China}
\date{\today}
\begin{abstract}
We have proposed the neck linker swing model to investigate the
mechanism of mechanochemical coupling of kinesin. The
Michaelis-Menten-like curve for velocity vs ATP concentration at
different loads has been obtained, which is in agreement with
experiments. We have predicted that Michaelis constant doesn't
increase monotonically and an elastic instability will happen with
increasing of applied force.
\end{abstract}

\maketitle
\section{introduction}
Kinesin is a molecular motor that transports organelles and
membrane-bound vesicles along a microtubule in various cells
\cite{val99,hir98}. It takes hundreds of $8\textsf{ nm}$ steps
(the size of tubulin heterodimers composed of $\alpha$ and $\beta$
subunits)\cite{blo90,svo93,how96} before detachment and the run
length is longer than $1\textsf{
$\mu$m}$\cite{blo90,svo93,how96,kaw00}. This is why kinesin is
called a processive motor.

Conventional kinesin (hereafter called `kinesin') is a dimer
consisting of two identical $\sim120$kD chains, commonly known as
heavy chains. Each heavy chain contains a N-terminal globular head
domain, a stalk region which is responsible for heavy chain
dimerization, a tether that joins the head and stalk, and a
C-terminal fan-shaped tail domain which usually binds cargo for
transporting in living cell or bead for applying force in single
molecule manipulated experiments \cite{hir99,hir89}. The head
domain is a highly conserved region for different members of the
kinesin superfamily and contains two binding sites. One binds to
microtubule and the other to nucleotides. Two heads alternately
hydrolyze one molecular of ATP for each $8\textsf{ nm}$
step\cite{val99,hir98,blo90,svo93,how96,kaw00,how89,hac95,hua97,sch97,coy99}.
Sometimes, the 8-$\textsf{nm}$ step can be resolved into fast and
slow substeps, each corresponding to a displacement of $\sim
4\textsf{ nm}$\cite{nis01}. The tether is a $\sim15$-amino-acid
segment. It becomes immobilized and extended towards the forward
direction when its head binds microtubule and ATP, and reverts to
a more mobile conformation when phosphate is released after ATP
hydrolysis\cite{ric99}. If the tether is replaced by a random
sequence of amino acids\cite{cas00} or if it is cross-linked to
nucleotides site\cite{tom00}, the kinesin will lose the capability
of stepping. Therefore, the conformational change of tether seems
to be necessary for kinesin to step\cite{ric99}.

The motility of kinesin can be explained by a ``hand over hand''
model\cite{kas03,asb03,yil04,sch04}. The two heads alternately
repeat single- and double-headed binding with microtubule. A
simplified binding mode has been proposed\cite{kaw01}. For
single-headed binding, the attached head either binds ATP or is
empty, whilst the detached head binds ADP. For double-headed
binding, the forward head is empty, whilst the rear head binds
either ATP or ADP$\cdot$P$_{\textrm{\footnotesize i}}$.

Kinesin works in a cyclic fashion for several intermediate states.
The Michaelis-Menten relation for the rate of ATP hydrolysis is
still a basic law\cite{how89,svo93,hac95,gil95,gil98,moy98,shu04}.
The average stepping velocity, however, also has been
experimentally found obeying the Michaelis-Menten law for a range
loads\cite{vis99}, which means the mechanochemical coupling is
tight, i.e., kinesin hydrolyzes one molecular of ATP for each
$8\textsf{ nm}$ step\cite{sch97}. With increasing of the applied
force, the saturating velocity decreases as expected, however,
Michaelis constant surprisingly increases\cite{vis99}. The role of
force in the reaction kinetics can be used to investigate the
mechanism of mechanochemical coupling.

In this paper, we present a neck linker swing model to investigate
the mechanism of mechanochemical coupling of kinesin. We will
discuss the effect of applied force on the reaction kinetics and
the elastic instability of the complex composed of neck linker and
attached head.

\section{neck linker swing model}
When a head attaches microtubule, its tether becomes a rigid rod
that can be bent by an applied force, and we call this tether as a
neck linker. Otherwise, we call the tether as a flexible chain if
its head is detached\cite{ric99}.

\begin{center}
\begin{figure}
\includegraphics[width=8.0 cm]{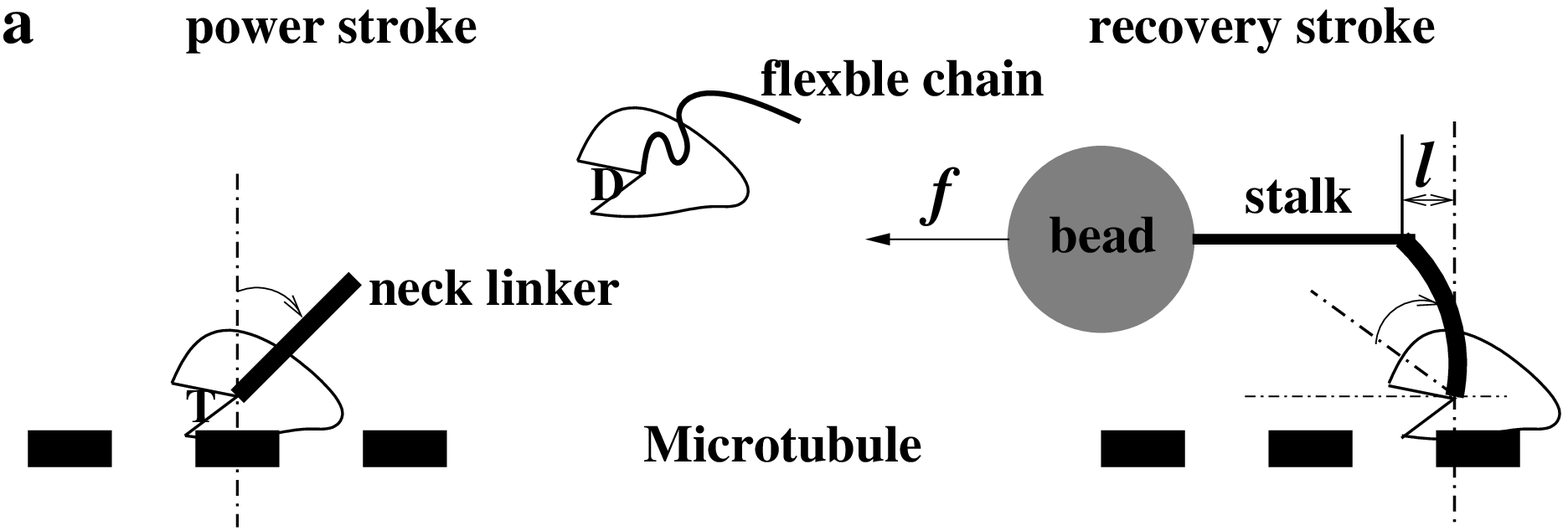}
\includegraphics[width=8.0 cm]{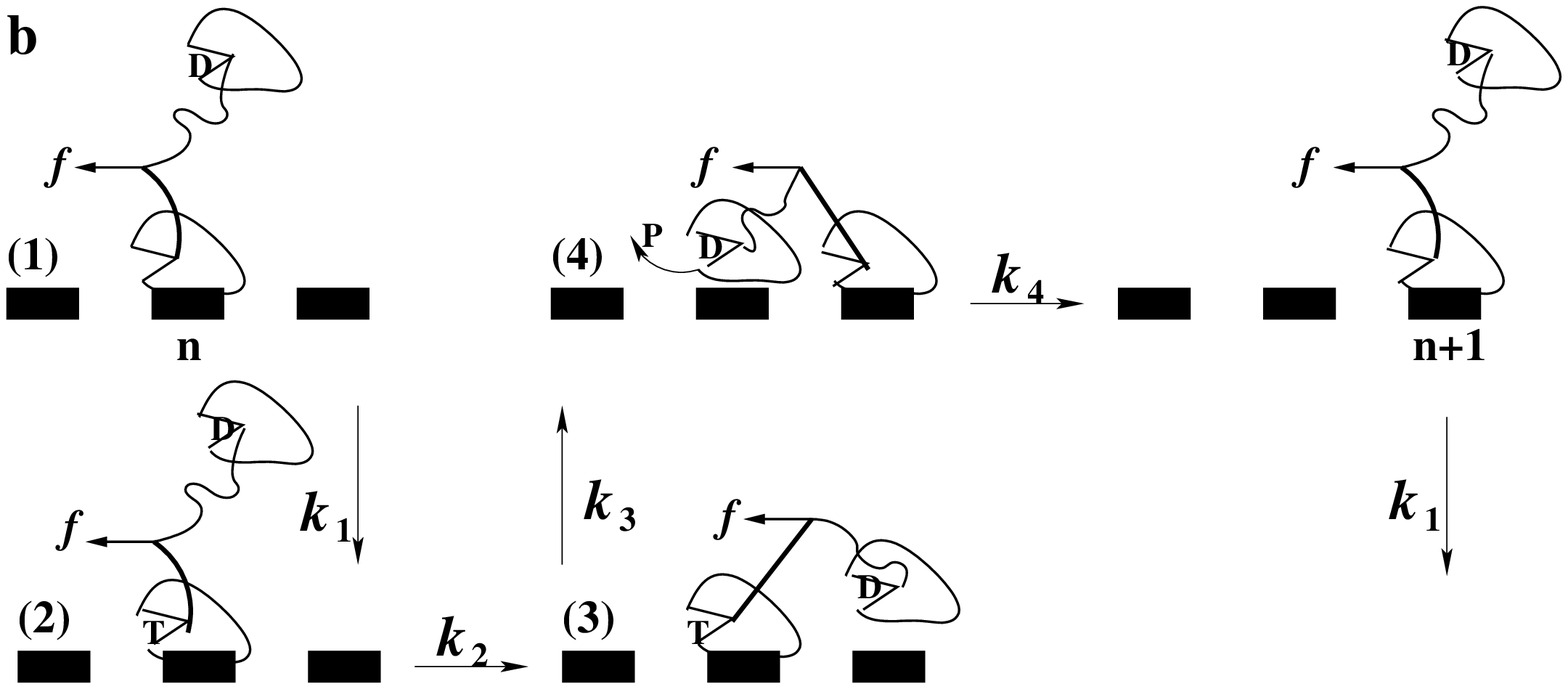}
\caption{(a): An idealized scheme of the relations among
nucleotides, microtubule and the tether's conformation(not to
scale). (b): Neck linker swing model. Two chemical transitions and
two mechanical substeps are coupled alternately to each other. T,
D and P represents ATP, ADP and P$_{\textrm{\footnotesize i}}$
respectively.} \label{f:neck}
\end{figure}
\end{center}

An idealized scheme of the relations among nucleotides,
microtubule and the tether's conformation is shown in Figure
\ref{f:neck}\textbf{a}\cite{ric99,val00,sos01,nel04}. For attached
head, if its catalytic cleft is empty, the neck linker tends to be
perpendicular to microtubule. ATP binding will change the
catalytic cleft's conformation. The allosteric interaction between
two binding sites in the attached head will trigger neck linker to
swing to the forward direction\cite{ric99}.

The neck linker swing model consists of two chemical transitions
($k_1$ and $k_3$) and two mechanical substeps ($k_2$ and $k_4$) as
shown in Figure \ref{f:neck}\textbf{b}. This model is consistent
with the widely accepted model for the kinetic mechanism of
kinesin\cite{cro04}. We just keep their main processes and
rearrange them into steps in our model. This approximation is
reasonable for the case that ADP concentration is low.

In this model, the mechanochemical cycle of kinesin includes four
steps: ATP binding; power stroke; ADP releasing + ATP hydrolysis
and recovery stroke.
\subsection{chemical transition 1: ATP binding}
When a head strongly attaches to the microtubule and waits for ATP
binding, the neck linker will be bent by the applied force $f$ to
the backward direction with displacement $\ell$ as shown in Figure
\ref{f:neck}\textbf{a}. The bending energy
$E_{\textrm{\footnotesize bend}}\approx\frac{1}{2}\kappa\ell^2$
and the force $ f={\rm d}E_{\textrm{\footnotesize bend}}/{\rm
d}\ell\approx\kappa\ell$ in the case of slight bending, where
$\kappa$ is the bending rigidity.

The attached head will finally bear the bending energy and the
energy barrier for ATP binding will increase. So ATP binding rate
can be written as:
\begin{eqnarray}
k_1(f)=k_{\textrm{\footnotesize b}}e^{-E_{\textrm{\tiny
bend}}/k_{\textrm{\tiny B}}T}\textrm{[ATP]}\label{eq:k1}
\end{eqnarray}
where $k_{\textrm{\footnotesize b}}$ is ATP binding constant
without applied force, $k_{\textrm{\tiny B}}$ is Boltzmann
constant, and $T$ is absolute temperature. This relationship holds
whether [ATP](ATP concentration) is high or low. During attached
head is waiting for ATP binding, the free head can't reach any
binding site on the track because the end-to-end distance of the
flexible chain is not long enough. These sites are separated by a
fixed distance ($L=8\textsf{ nm}$) along the rigid microtubule.
\subsection{mechanical substep 1: power stroke}
Once the attached head binds an ATP molecule, neck linker will
move to the new equilibrium position, and the free head and bead
will be thrown forward. A power stroke occurs. We assume the
motion of bead, by which optical tweezer can applies force on
motor, is overdamped, the average velocity is
\begin{eqnarray}
\langle\dot{x}\rangle=\frac{1}{\zeta}(\langle
f_{\textrm{\footnotesize c}}\rangle+\langle
f_{\textrm{\footnotesize b}}\rangle -f)\label{eq:v}
\end{eqnarray}
where $\zeta$ is the viscous coefficient, $\langle
f_{\textrm{\footnotesize c}}\rangle$ and $\langle
f_{\textrm{\footnotesize b}}\rangle$ are average forces generated
from ATP binding and releasing of the bending energy respectively,
that is, $\langle f_{\textrm{\footnotesize c}}\rangle+\langle
f_{\textrm{\footnotesize b}}\rangle=(\Delta
E_1+E_{\textrm{\footnotesize bend}})/(\frac{1}{2}L+\ell)$. The
power stroke rate then is
\begin{eqnarray}
k_2(f)=\frac{\langle
\dot{x}\rangle}{L/2+\ell}=\frac{1/\zeta}{L/2+\ell}\left(\frac{\Delta
E_1+E_{\textrm{\footnotesize
bend}}}{L/2+\ell}-f\right)\label{eq:k2}
\end{eqnarray}
$\Delta E_1$ is the energy that power stroke outputs, and
originates from the catalytic cleft's conformational change
induced by ATP binding\cite{ric99}.
\subsection{chemical transition 2: detached head's ADP releasing and attached head's ATP hydrolysis}\label{s2.2}
With starting of the power stroke, the random motion of the free
head will be biased to the forward direction and the flexible
chain will be stretched due to neck linker's throwing and the
interaction from the nearest $\beta$-tubulin monomer, which
greatly increases the probability that the free head reaches the
next binding site on the track. It first bind weakly, eventually,
releases its ADP and attaches strongly to the track. The stretched
chain will shrink and transform into a rigid rod, which helps the
rear head to split the bound ATP, and one of products, phosphate
will be released\cite{ric99,kaw01,nel04}. This reaction weakens
the binding of the rear head with the microtubule and leads it to
detach from the track with ADP\cite{ric99,kaw01,nel04}. We assume
the rate of the process from the end of power stroke to the
detaching of rear head, $k_3$, is independent of applied force.
\subsection{mechanical substep 2: recovery stroke}
The tether linked the forward head now becomes the neck
linker\cite{ric99,kaw01,nel04}, and will swing to the mechanical
equilibrium position. A recovery stroke occurs. Similar with the
power stroke, the rate of this step
\begin{eqnarray}
k_4(f)=\frac{1/\zeta}{L/2-\ell}\left(\frac{\Delta
E_2-E_{\textrm{\footnotesize
bend}}}{L/2-\ell}-f\right)\label{eq:k4}
\end{eqnarray}
where $\Delta E_2$ originates from the catalytic cleft's
conformational change induced by ADP releasing, which is
conformationally the recovery process of ATP binding.

The cycle is now ready to repeat, the difference is that roles of
these two partner heads have exchanged. The kinesin dimer has
hydrolyzed one ATP and moved forward $8\textsf{ nm}$ with two
substeps\cite{ric99,val00,sos01,nel04}. Although there are more
than two chemical transitions as above, all other chemical
processes are rapid rate transitions (randomness shows that there
are only two to three rate-limiting
transitions\cite{vis99,blo03,sch95}). We lump all other chemical
transitions into $k_3$.

\section{fitting and results}\label{s3}

From the time series shown in Figure\ref{f:neck}\textbf{b}, the
average time to complete a single enzymatic cycle,
$\langle\tau\rangle$, can be computed
conveniently\cite{sch95,sun01}. The average velocity of kinesin
moving along microtubule is
\begin{eqnarray}
v=\frac{L}{\langle\tau\rangle}=L\left(\sum_{i=1}^4\frac{1}{k_i}\right)^{-1}
=\frac{v_{\textrm{\footnotesize
max}}\textrm{[ATP]}}{K_{\textrm{\footnotesize
M}}+\textrm{[ATP]}}\label{eq:v1}
\end{eqnarray}
with
\begin{eqnarray}
v_{\textrm{\footnotesize max}}&=&L\left(\frac{1}{k_2}+\frac{1}{k_3}+\frac{1}{k_4}\right)^{-1}\label{eq:vm}\\
K_{\textrm{\footnotesize M}}&=&\frac{e^{E_{\textrm{\tiny
bend}}/k_{\textrm{\tiny B}}T}}{k_{\textrm{\footnotesize
b}}}\frac{v_{\textrm{\footnotesize max}}}{L}\label{eq:km}
\end{eqnarray}
where $v_{\textrm{\footnotesize max}}$ and
$K_{\textrm{\footnotesize M}}$ are saturating velocity and
Michaelis constant respectively. Obviously, the average velocity
obeys the Michaelis-Menten law as observed\cite{vis99}.

In our model, $\Delta E_1$ and $\Delta E_2$ are independent of ATP
concentration. We define the efficiency of mechanochemical
coupling of kinesin as $\eta=(\Delta E_1+\Delta E_2)/\Delta\mu$
and assume it is a constant, where $\Delta\mu$ is the free energy
excess in one molecule ATP hydrolysis. It must be noted that
$\eta$ defined here is different from the efficiency of motor. We
assume kinesin has adjusted $\Delta E_1$ and $\Delta E_2$ to
achieve optimal kinetic velocity in evolution. From
Eq.(\ref{eq:k2}) and Eq.(\ref{eq:k4}), we yield $\Delta E_1=\Delta
E_2\equiv\Delta E$ ($<\frac{1}{2}\Delta\mu$), which is reasonable
with the fact that the catalytic cleft's conformational change
induced by ADP releasing is conformationally the recovery process
of that induced by ATP binding. For simplicity, we introduce the
rate of power (or recovery) stroke without load, $k_0=4\Delta
E/(\zeta L^2)$.

According to Eqs.(\ref{eq:k2}) and (\ref{eq:k4}), the power stroke
rate $k_2$ decreases with increasing of applied force $f$, whilst
the recovery stroke rate $k_4$ increases as shown in
Figure\ref{f:f2}\textbf{(d)}. At a critical applied force, $k_2$
becomes zero. We call this critical force as stall force. It can
be derived as
\begin{eqnarray}
f_{\textrm{\tiny stall}}=\frac{\Delta E+E_{\textrm{\footnotesize
bend}}(f_{\textrm{\tiny stall}})}{L/2+\ell(f_{\textrm{\tiny
stall}})}\label{eq:f1}
\end{eqnarray}
If applied force is larger than the stall force, the force induced
by conformation change can't overcome it to push the kinesin
forward along the microtubule.

For the case of slight bending of neck linker, Eq.(\ref{eq:f1})
can be rewritten as $f_{\textrm{\tiny stall}}\approx2\Delta
E/[L(1+f_{\textrm{\tiny stall}}/(\kappa
L))]<\Delta\mu/L\sim10\textsf{ pN}$ with
$\Delta\mu\approx80\textsf{ pN$\cdot$nm}$ under physiological
conditions\cite{str95}, which is in agreement with the
experimental data\cite{vis99}. It must be noted that stall force
in our model is independent of ATP concentration. We can rewrite
Eq.(\ref{eq:vm}) and Eq.(\ref{eq:km}) as
\begin{eqnarray}
\ln\left(\frac{K_{\textrm{\footnotesize
M}}L}{v_{\textrm{\footnotesize max}}}\right)=-\ln
k_{\textrm{\footnotesize b}}+\frac{1/(2k_{\textrm{\tiny
B}}T)}{\kappa}f^2\qquad\qquad\label{eq:ft1}\\
\frac{L}{v_{\textrm{\footnotesize
max}}}=\frac{1}{k_3}+\frac{1}{k_0}\left[\frac{(1+2\bar{f})^2}{1-\bar{f}_{\Delta}(1+\bar{f})}
+\frac{(1-2\bar{f})^2}{1-\bar{f}_{\Delta}(1-\bar{f})}\right]\label{eq:ft2}
\end{eqnarray}
and use them to fit the measured data of $v_{\textrm{\footnotesize
max}}$ and $K_{\textrm{\footnotesize M}}$ at different
loads\cite{vis99}, where $\bar{f}_{\Delta}=f/[f_{\textrm{\tiny
stall}}(1+f_{\textrm{\tiny stall}}/(\kappa L))]$, and
$\bar{f}=f/(\kappa L)$. The fitted values of $f_{\textrm{\tiny
stall}}$, $\kappa$, $k_{\textrm{\footnotesize b}}$, $k_3$ and
$\zeta$ are listed in table \ref{table1.1}, and the two fitted
curves, $v_{\textrm{\footnotesize max}}\sim f$ and
$K_{\textrm{\footnotesize M}}\sim f$, are shown in Figure
\ref{f:f2}\textbf{(a)}.

\begin{table}
\centering \caption{The fitted parameters for experimental
$v_{\textrm{\footnotesize max}}$ and $K_{\textrm{\footnotesize
M}}$  versus $f$\cite{vis99} with $k_{\textrm{\tiny
B}}T_{\textrm{\footnotesize r}}=4.1 \textsf{pN$\cdot$nm}$ ¡£}
\label{table1.1}
\tabcolsep0.25cm
\begin{tabular}{ccccc}
\hline $f_{\textrm{\tiny stall}}$ & $\kappa$&
$k_{\textrm{\footnotesize b}}$ & $k_3$& $\zeta$\\
(\textsf{pN}) & (\textsf{pN$\cdot$nm$^{-1}$})&
(\textsf{$\mu$M$^{-1}\cdot$s$^{-1}$})& (\textsf{s$^{-1}$})&
(\textsf{pN$\cdot$s/nm})\\
\hline \hline 6.50 & 1.88 & 1.33 & 120 &
1.45$\times10^{-3}$\\
\hline
\end{tabular}
\end{table}

The fitted stall force $f_{\textrm{\tiny stall}}=6.5\textsf{ pN}$
is consistent with measured value\cite{vis99}. The total energy
outputted in one cycle, $2\Delta E\approx74\textsf{ pN$\cdot$nm}$,
is less than $\Delta\mu$, which demonstrated again that the fitted
values are reasonable. The efficiency of mechanochemical coupling
of kinesin, $\eta$, is about 93\%. However, the efficiency of
motor, $f_{\textrm{\tiny stall}}L/\Delta\mu$, is about 65\%, which
is in agreement with experiments.

It is the motion of a silica bead that is measured in
experiment\cite{vis99}. The radius of the bead, $R$, is
$0.25\textsf{ $\mu$m}$. The viscousity of water at room
temperature, $\tilde{\eta}_w$, is about
$0.9\times10^{-9}\textsf{pN$\cdot$s/nm$^2$}$. The Stokes drag
coefficient of the bead can be estimated by $6\pi
R\tilde{\eta}_w$, and is about
$4.24\times10^{-3}\textsf{pN$\cdot$s/nm}$. The fitted viscous
coefficient of bead in our model, thus, is comparable with the
Stokes drag coefficient for a sphere in $0.5\textsf{ $\mu$m}$
diameter at room temperature.

The widely accepted model for the kinetic mechanism of
kinesin\cite{cro04} has proposed that the ATP binding rate is
about $2\textsf{ $\mu${\footnotesize M}$^{-1}\cdot$s$^{-1}$}$ and
the rate from free head's ADP releasing to attached head's
detaching can be estimated about $70\textsf{ s$^{-1}$}$. In our
model, ATP dissociating rate $k_{-1}$ isn't taken into account,
this is why $k_{\textrm{\footnotesize b}}$ is slightly less than
the proposed value. $k_3$ also approaches the estimated value.
\section{discussion}
\subsection{$K_{\textrm{\footnotesize M}}$ no longer increases monotonically with load}\label{s3.2}
\begin{center}
\begin{figure}
\includegraphics[width=8.0 cm]{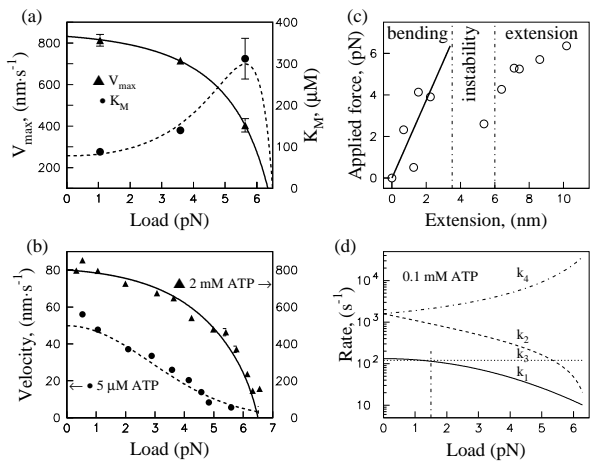}
\caption{\textbf{(a)}: Fitting to experimental
$v_{\textrm{\footnotesize max}}$ and $K_{\textrm{\footnotesize
M}}$ vs $f$\cite{vis99}. The fitted parameters are listed in table
\ref{table1.1}. \textbf{(b)}: Dependence of velocity on load at
saturated and limiting ATP concentrations. The curves are directly
from Eq.(\ref{eq:v1}) analytically with parameters shown in table
\ref{table1.1}, and the experimental data come from
Ref\cite{vis99}. \textbf{(c)}: Applied force vs extension. The
relation $f=\kappa\ell$ with $\kappa$ shown in table
\ref{table1.1} is in good agreement with the experimental
data\cite{kaw01} at low force. \textbf{(d)}: Dependence of rates
on applied force at $0.1\textsf{ m\footnotesize M}$ ATP.}
\label{f:f2}
\end{figure}
\end{center}

Although the recovery stroke will go faster and faster with
increasing of applied force, the power stroke will spend more and
more time and finally stalls at the stall force, which leads to
the concave down of saturating velocity. With low applied force,
Michaelis constant will nearly increase exponentially because
$e^{E_{\textrm{\tiny bend}}/k_{\textrm{\tiny B}}T}$ increases
faster than $v_{\textrm{\footnotesize max}}$ decreases. With high
force, power stroke becomes the rate-limiting step for the single
enzymatic cycle, so, $K_{\textrm{\footnotesize M}}$ will fall fast
as $v_{\textrm{\footnotesize max}}$ shown in
Figure\ref{f:f2}\textbf{(a)}. Michaelis constant no longer
increases monotonically with load. The big error bar at high force
implies that this prediction is possible.

\subsection{neck linker's bending}

Actually, the neck linker's bending has been unconsciously
revealed in experiment\cite{kaw01}. With an applied force moving
toward the end of microtubule, a simply elastic model can fit all
extension of kinesin-microtubule complex for double-headed
attaching. But the force-extension relation in
Figure\ref{f:f2}\textbf{(c)} can't be fitted by a same simply
elastic model for all extension for single-headed attaching. With
our neck linker swing model, we can image there are three
distinctive regions of force. If the force is low, the
``Extension'' is mainly contributed by neck linker bending. We
compare the relation $f=\kappa\ell$ to the experimental
force-extension data\cite{kaw01} in Figure \ref{f:f2}\textbf{(c)}.
It is very surprising that our model is in agreement with
experiment very well. With the increasing of applied force, a
phenomenon of elastic instability\cite{lan86} will inevitably
happen to the complex composed of neck linker and attached head.
This is why the extension rapidly increases while the ``applied
force'' falls fast as the measured data in Figure
\ref{f:f2}\textbf{(c)}. After reaching the new mechanical
equilibrium, the complex will further extend with increasing of
applied force, and its force-extension relation can be fitted by a
simply elastic model again.

\subsection{velocity versus applied force at saturating and limiting ATP concentration}

With these reasonable fitted parameters, we can use
Eq.(\ref{eq:v1}) directly to compute the dependence of the
velocity on load at saturating and limiting ATP concentrations. As
shown in Figure \ref{f:f2}\textbf{(b)}, the theoretical
velocity-force relation is in good agreement to experimental
data\cite{vis99}. It is clear that there are three distinctive
regimes of applied force: (1) If the load is low, chemical
transition 2, the process from free head's ADP releasing to
attached head's ATP hydrolysis, $k_3$, is the rate-limiting
transition at saturating [ATP] and $v\approx
Lk_3\approx900\textsf{ nm/s}$, which is consistent with what is
known about the biochemistry of kinesin\cite{hac88,gil94}, while
ATP binding is the rate-limiting transition at low [ATP] and
$v\approx Lk_1\approx50\textsf{ nm/s}$ with $5\textsf{
$\mu$\footnotesize M}$. (2) If load is high, kinesin will stall at
the same force as discussed in section \ref{s3} whether ATP
concentration is high or low. (3) If applied force is moderate,
the velocity-force curves display different shapes at low and
saturating ATP concentration respectively. At very low ATP
concentration, the velocity decreases exponentially as $k_1$ and
looks like linear with load because ATP binding is the
rate-limiting step. At saturating [ATP], the velocity is concave
down with load as discussed in section \ref{s3.2}.

\subsection{two mechanical substeps}

The two mechanical substeps in this model are contributed by the
two heads respectively. The recovery stroke is always a rapid rate
step which corresponds to the observed fast substep\cite{nis01}.
If a moderate load such as $1.5\textsf{ pN}$ acts on the bead and
ATP concentration is maintained at $0.1\textsf{ m\footnotesize
M}$, the two chemical transitions have the same rate as shown in
Figure\ref{f:f2}\textbf{(d)}. The time spent in ATP binding equals
that spent in chemical transition 2. Theoretically, $4.8\textsf{
nm}$ slow substep and $3.2\textsf{ nm}$ fast substep can be
detected directly by single molecular manipulated techniques such
as optical tweezers. The different load-dependence of these
substeps' rate may be revealed in the future experiment.

\section{conclusion}
We proposed the neck linker swing model which divides the single
enzymatic cycle into two chemical transitions and two mechanical
substeps. Each chemical transition will induce the conformational
change in the catalytic cleft and generate a corresponded
mechanical stroke. The model can be used to explain the observed
substeps\cite{nis01}. The different load-dependence of these two
strokes' rate may be revealed in the future experiment. We have
investigated the mechanism of mechanochemical coupling of kinesin
by the influence of applied force on the bending of neck linker.
When attached head is waiting for ATP binding, the neck linker is
bent by the applied force. The attached head bears the neck
linker's bending energy and the energy barrier for ATP binding
increases. This is why Michaelis constant increases with applied
force. Our theoretical analysis of average velocity of motor in
Eq.(\ref{eq:v1}) also obeys Michaelis-Menten law and has been used
to fit the observed saturating velocity and Michaelis constant at
different loads\cite{vis99}. The fitted values of chemical
reaction rates are in agreement with those in the widely accepted
model\cite{cro04}, and the fitted viscous coefficient of bead is
also comparable with the Stokes drag coefficient. The stall force
is independent of ATP concentration and its fitted value is
consistent with the observed data in experiment\cite{vis99}. The
fitted bending rigidity of neck linker can be used to explain the
relation of force-extension in experiment at low
force\cite{kaw01}. With these reasonable fitted parameters, we can
directly use Eq.(\ref{eq:v1}) to describe the relation between the
average velocity and load at different ATP concentrations, which
is in good agreement to experimental data\cite{vis99} as shown in
Figure \ref{f:f2}\textbf{(b)}. In addition, we have predicted
Michaelis constant doesn't increase monotonically and an elastic
instability will happen to the complex composed of neck linker and
attached head with increasing of applied force.

\section*{Acknowledgements} We acknowledge useful discussions with
Ou-Yang Zhong-can and Ming Li. This work was supported by Special
Fund for Theoretical Physics of Postdoctor and National Science
Foundation of China.


\begin{thebibliography}{99}
\bibitem{val99} T.Kreis and R.D.Vale , \textit{Guidebook to the Cytoskeletal and Motor Proteins},
(Oxford Univ. Press, Oxford, ed. 2, 1999) pp.398-402
\bibitem{hir98} N.Hirokawa, Science, {\bf 279}, 519 (1998)
\bibitem{blo90} S.M.Block, L.S.B.Goldstein and B.J.Schnapp, Nature, {\bf 348}, 348 (1990)
\bibitem{svo93} K.Svoboda \textit{et al.,} Nature {\bf 365}, 721 (1993)
\bibitem{how96} J.Howard, Annu. Rev. Physiol. {\bf 58}, 703 (1996)
\bibitem{kaw00} K.Kawaguchi and S.Ishiwata, Biochem. Biophys. Res. Commun. {\bf 272}, 895 (2000)
\bibitem{hir99} K.Hirose, L.A.Amos, Cell. Mol. Life Sci. {\bf 56}, 184(1999)
\bibitem{hir89} N.Hirokawa \textit{et al.,} Cell {\bf 56}, 867(1989)
\bibitem{how89} J.Howard, A.J.Hudspeth and R.D.Vale, Nature {\bf 342}, 154 (1989)
\bibitem{hac95} D.D.Hackney, Nature {\bf 377}, 448 (1995)
\bibitem{hua97} W.Hua \textit{et al.,} Nature {\bf 388}, 390 (1997)
\bibitem{sch97} M.J.Schnitzer and S.M.Block, Nature {\bf 388}, 386 (1997)
\bibitem{coy99} D.L.Coy, M.Wagenbach and J.Howard, J. Biol. Chem. {\bf 274}, 3667 (1999)
\bibitem{nis01} M.Nishiyama \textit{et al.,} Nat. Cell Biol., {\bf 3}, 425 (2001)
\bibitem{ric99} S.Rice \textit{et al.}, Nature {\bf 402}, 778 (1999)
\bibitem{cas00} R.B.Case \textit{et al.,} Curr. Biol., {\bf 10}, 157 (2000)
\bibitem{tom00} M.Tomishige and R.D.Vale, J. Cell. Biol. {\bf 151}, 1081 (2000)
\bibitem{kas03} K.Kasedal, H.Higuchi and K.Hirosel, Nat. Cell Biol., {\bf 5}, 1079 (2003)
\bibitem{asb03} C.L.Asbury, A.N.Fehr and S.M.Block, Science, {\bf 302}, 2130 (2003)
\bibitem{yil04} A.Yildiz \textit{et al.,} Science, {\bf 303}, 676 (2004)
\bibitem{sch04} W.R.Schief \textit{et al.,} PNAS, {\bf 101}, 1183 (2004)
\bibitem{kaw01} K.Kawaguchi and S.Ishiwata, Science, {\bf 291}, 667 (2001)
\bibitem{gil95} S.P.Gilbert \textit{et al.,} Nature, {\bf 373}, 671 (1995)
\bibitem{gil98} S.P.Gilbert, M.L.Moyer, K.A.Johnson, Biochemistry, {\bf 37}, 792 (1998)
\bibitem{moy98} M.L.Moyer, S.P.Gilbert, K.A.Johnson, Biochemistry, {\bf 37}, 800 (1998)
\bibitem{shu04} Y.G.Shu and H.L.Shi, PRE, {\bf 69}, 021912 (2004)
\bibitem{vis99} K.Visscher, M.J.Schnitzer and S.M.Block, Nature {\bf 400}, 184 (1999)
\bibitem{val00} R.D.Vale and R.A.Milligan, Science {\bf 288}, 88 (2000)
\bibitem{sos01} H.Sosa \textit{et al.,} Nat. Struct. Biol., {\bf 8}, 540 (2001)
\bibitem{nel04} P.Nelson, \textit{Biological Physics: Energy, Information, Life}, (W.H. Freeman and Co., 2004),pp.441-444.
\bibitem{cro04} R.A.Cross, TRENDS in Biochemical Sciences, {\bf 29}, 301 (2004)
\bibitem{blo03} S.M.Block \textit{et al.,} PNAS {\bf 100}, 2351 (2003)
\bibitem{sch95} M.J.Schnitzer and S.M.Block, Cold Spring Harbor Symposia on Quantitative Biology, {\bf 60}, 793 (1995)
\bibitem{sun01} Sunney Xie, Single Mol. {\bf 2}, 229 (2001)
\bibitem{str95} L.Stryer, \textit{Biochemistry}, Fourth Edition (New York: Freeman,1995),pp.443-462.
\bibitem{lan86} L.D.Landau and E.M.Lifshitz, \textit{Theory of Elasticity}, (Peframon Press, 1986), pp.70-84.
\bibitem{hac88} D.D.Hackney, PNAS, {\bf 85}, 6314 (1988)
\bibitem{gil94} S.P.Gilbert and K.A.Johnson, Biochemistry, {\bf 33}, 1951 (1994)
\end{thebibliography}
\end{document}